\documentclass[a4paper,10pt]{article}
\usepackage{moreverb}
\usepackage{natbib}
\usepackage{times,epsfig,makeidx,amsfonts,amsmath,dcolumn,enumerate,multirow,graphicx,amssymb,colortbl,bm,url,hyperref}
\usepackage{algorithm}
\usepackage[]{algorithmic}
\usepackage{mathtools,upgreek}
\usepackage{afterpage,lscape}
\usepackage{paralist}
\usepackage{calrsfs}
\usepackage[margin=1in]{geometry}

\usepackage{epstopdf}

\DeclareMathAlphabet{\pazocal}{OMS}{zplm}{m}{n}

\newcommand{\bx}{{\bf x}}

\newcommand{\bz}{{\bf z}}

\begin{document}

\title{On a general structure for hazard-based regression models: an application to population-based cancer research}

\author{Francisco J. Rubio$^a$$^*$, Laurent Remontet$^b$, Nicholas P. Jewell$^c$,  Aur{\'e}lien Belot$^a$
\\ \small $^{a}$ Cancer Survival Group, Faculty of Epidemiology and Population Health,
\\ \small Department of Non-Communicable Disease Epidemiology,
\\ \small London School of Hygiene \& Tropical Medicine,
\\ \small London, WC1E 7HT, United Kingdom.
\\ \small $^{b}$ Hospices Civils de Lyon,
\\ \small Service de Biostatistique-Bioinformatique,
\\ \small Lyon, 69003, France.
\\ \small $^{c}$ University of California, Berkeley,
\\ \small School of Public Health,
\\ \small Berkeley, CA 94720-7360.
\\ \small $^{*}$ E-mail:  Francisco.Rubio@lshtm.ac.uk}

\maketitle

\begin{abstract}
The proportional hazards model represents the most commonly assumed hazard structure
when analysing time to event data using regression models. We study a general hazard structure which contains, as particular cases, proportional hazards, accelerated hazards, and accelerated failure time structures, as well as combinations of these. We propose an approach to apply these different hazard structures, based on a flexible parametric distribution (Exponentiated Weibull) for the baseline hazard. This distribution allows us to cover the basic hazard shapes of interest in practice: constant, bathtub, increasing, decreasing, and unimodal. In an extensive simulation study, we evaluate our approach in the context of excess hazard modelling, which is the main quantity of interest in descriptive cancer epidemiology. This study exhibits good inferential properties of the proposed model, as well as good performance when using the Akaike Information Criterion for selecting the hazard structure. An application on lung cancer data illustrates the usefulness of the proposed model.\\

\end{abstract}

\section{Introduction}\label{sec:intro}
The analysis of time-to-event data has been dominated by the use of the semi-parametric Cox model during the last decades. While extensions to remove the assumption of \textit{proportional hazards} (PH) were already discussed in Cox's original paper, a lot of work has been devoted to improve the flexibility of hazard-based regression models using flexible functions for both the baseline hazard and the inclusion of time-dependent parameters, mainly using splines or fractional polynomials \citep{Durrleman89, Sleeper90, Hess94, Abrahamowicz96, royston2008multivariable}. However, the general structure of those models remained the same: the hazard function was expressed with a baseline hazard $h_0$ multiplied by the exponential of a flexible function $g$ of time $t$ and covariates $\bx$:  $h(t;\bx)=h_0(t)\exp(g(t;\bx))$. Alternative hazard structures that directly account for time-dependent effects of covariates have been proposed; one of the earliest alternative to the PH model is the \textit{accelerated failure time} (AFT) model, in which the variables have a direct effect on the time to event, in contrast to the PH model where the covariates affect the hazard function \citep{K11}. More recently, in a series of papers \citep{C00,C01,C03,C14}, Y.~Chen and co-authors proposed and studied semiparametric \textit{accelerated hazards} (AH) models, where the covariates have a time-scaling effect on the hazard function, thus allowing for a time-varying effect.

In cancer epidemiology using population-based registry data, the quantity of interest is usually the excess hazard (of death) instead of the overall hazard \citep{H87, E90, R07, M14}. The basic idea behind excess hazard models consists of decomposing the hazard function of death associated with an individual, $h(t;\bx)$, as the sum of an excess hazard, $h_{\text{E}}(t;\bx)$, and the general population hazard $h_{\text{P}}(t;\bz)$. The general population hazard is supposed to correctly reflect the other-causes hazard of death in our population of interest, assuming that the contribution in the general population hazard of the disease of interest (\textit{e.g.}~a specific cancer type) is small compared to all others. This excess hazard is interpreted as the hazard that could be due, directly or indirectly, to the cancer under study. Formally, this can be written as:
\begin{eqnarray}\label{HazardAddModel}
h(t;\bx) = h_{\text{P}}(\text{age}+t; \text{year}+t;\bz) + h_{\text{E}}(t;\bx),
\end{eqnarray}
where $\bx$ and $\bz$ are vectors of covariates, $\bz$ typically corresponding to a subset of covariates of $\bx$, ``age'' is the age at diagnosis and ``year'' is the year of diagnosis (thus ``$\text{age}+t$'' and ``$\text{year}+t$'' are respectively the age and the year at time $t$ after diagnosis). The population hazard $h_{\text{P}}(\cdot;\bz)$ is typically obtained from national life tables based on the available sociodemographic characteristics ($\bz$ in addition to age and year).
A survival function derived from the excess hazard $h_{\text{E}}(\cdot;\bx)$ is called the \emph{net survival}.  Net survival represents a useful way of reporting the probability of survival of cancer patients since this allows for a fairer comparison of survival rates between different populations or countries \citep{GV01, PS12, D12, P16}, and a non-parametric estimator of net survival has been proposed recently \citep{PS12}. Although the interest in cancer epidemiology is on a slightly different quantity (the excess hazard instead of the overall hazard), the methodological developments of regression models have followed the same path than those described above \citep{B02, G03, C07, R07, M11, C16} in terms of the hazard structure adopted.

Our aim is to provide a valuable supplement in the available toolbox for analysing survival data, and which is applicable for both overall and excess hazard regression models. The proposed approach builds on top of two recent developments: (i) the general parametrisation of hazard functions \citep{C00,C01}, combined with (ii) the use of a flexible parametric distribution for modelling times to event, the exponentiated Weibull distribution \citep{M96}.
The paper is organised as follows. In Section \ref{sec:Methods}, we describe the general hazard structure, discuss how the accelerated failure time model, the proportional hazards model, and the accelerated hazards model are nested within this general structure, and discuss the parameter interpretations for these models. In this section, we also introduce the exponentiated Weibull distribution, which will be used to model the ``baseline'' hazard in the general hazard structure (and the corresponding sub-models) based on its flexibility and ease of implementation. We also discuss maximum likelihood estimation and inference associated with these models. In Section \ref{sec:simulation}, we present an extensive simulation study where we illustrate the inferential properties of the proposed models. In Section \ref{sec:Example}, we present a data example from lung cancer epidemiology. Here, we illustrate the usefulness of the proposed models and compare them against appropriate competitor approaches. We conclude with a general discussion and point to possible extensions in Section \ref{sec:discussion}. Additional simulations and results are provided in the Supplementary Material.


\section{Methods}\label{sec:Methods}

\subsection{The different model structures}\label{sec:Model}
We considered the following excess hazard structures (see \cite{C00}, \cite{C01} and \cite{C03} for extensive discussion). We express the different structures below via the hazard function $h()$ and the cumulative hazard function $H()$, respectively, according to time $t$ and a vector of covariates $\bx_j$. We assume that the vector of covariates does not contain an intercept, in order to avoid identifiability issues. The survival function can be obtained from the well-known relationship $S(t) = \exp[-H(t)]$. The vector $\beta$ denotes the unknown regression parameters.
\begin{enumerate}[(i)]
\item Proportional hazards model (PH).
\begin{eqnarray}\label{PH}
h_{\text{E}}^{\text{PH}}\left(t;\bx_j\right) &=& h_0(t)\exp\left(\bx_j^T\beta\right),\\
H_{\text{E}}^{\text{PH}}\left(t;\bx_j\right) &=& H_0(t)\exp\left(\bx_j^T\beta\right).\nonumber
\end{eqnarray}
\item Accelerated hazards model (AH).
\begin{eqnarray}\label{AH}
h_{\text{E}}^{\text{AH}}\left(t;\bx_j\right) &=& h_0\left(t \exp(\bx_j^T\beta)\right),\\
H_{\text{E}}^{\text{AH}}\left(t;\bx_j\right) &=& H_0\left(t \exp(\bx_j^T\beta)\right)\exp(-\bx_j^T\beta).\nonumber
\end{eqnarray}
\item Hybrid hazards model (HH).
\begin{eqnarray}\label{HH}
h_{\text{E}}^{\text{HH}}\left(t;\bx_j\right) &=& h_0\left(t \exp(\bx_{1j}^T\beta_1)\right)\exp(\bx_{2j}^T\beta_2),\\
H_{\text{E}}^{\text{HH}}\left(t;\bx_j\right) &=& H_0\left(t \exp(\bx_{1j}^T\beta_1)\right)\exp(-\bx_{1j}^T\beta_1+\bx_{2j}^T\beta_2),\nonumber
\end{eqnarray}
where  $\bx_{1j} \subseteq \bx_j$, and $\bx_{2j} \subseteq \bx_{j}$.
\item Accelerated failure time model (AFT).
\begin{eqnarray}\label{AFT}
h_{\text{E}}^{\text{AFT}}\left(t;\bx_j\right) &=& h_0\left(t \exp(\bx_j^T\beta)\right)\exp(\bx_j^T\beta),\\
H_{\text{E}}^{\text{AFT}}\left(t;\bx_j\right) &=& H_0\left(t \exp(\bx_j^T\beta)\right).\nonumber
\end{eqnarray}
\item General hazards model (GH).
\begin{eqnarray}\label{GH}
h_{\text{E}}^{\text{GH}}\left(t;\bx_j\right) &=& h_0\left(t \exp(\bx_j^T\beta_1)\right)\exp(\bx_j^T\beta_2),\\
H_{\text{E}}^{\text{GH}}\left(t;\bx_j\right) &=& H_0\left(t \exp(\bx_j^T\beta_1)\right)\exp(-\bx_j^T\beta_1+\bx_j^T\beta_2).\nonumber
\end{eqnarray}
\end{enumerate}
The assumptions behind each of these models are different in nature. The basic idea is to include effects that affect the level of the hazard (time-fixed effects) and the time scale (time-dependent effects) separately, as follows. In the PH model \eqref{PH}, a unit change in a covariate value have a multiplicative effect on the hazard, thus leading to a level change on the y-axis of the hazard. In the AH model \eqref{AH}, the effect of a unit change in a covariate affects the time scale of the baseline hazard (x-axis), thus assuming that there is a time-dependent effect of each covariate. In other words, $\exp(\beta)$ could be seen as a factor of how much more (or less) time is needed to reach the same hazard level, as compared to baseline, when the covariate is increased by one unit. In addition, the AH model does not assume that the parameters affect the hazard immediately at time $t=0$, but ``gradually'', which is not the case with the other hazard structures \cite{C14}. While a ``gradual'' effect can be useful when estimating a treatment effect, it may not be justified for other variables; it highlights the usefulness of the HH structure \eqref{HH}. Indeed, the HH relaxes the assumption of the AH model by allowing some variables to have a proportional hazards effect rather than time-dependent ``gradual'' effects. In the AFT model \eqref{AFT}, the effect is on the survival time directly as it can be, in fact, formulated as a log-linear regression model on survival times \citep{K11}. In such models, the estimated regression parameter for a unit change in one covariate accelerates or decelerates the event time, thus leading to a time-dependent effect. Notice that the AFT, PH, and AH models coincide for the case when the baseline hazard is Weibull \citep{C01}. The GH model \eqref{GH} represents a general hazard structure that contains, as particular cases, the PH, AH, HH, and AFT models. More specifically, if $\beta_1=0$, then $\text{GH}=\text{PH}$; if $\beta_2=0$, then $\text{GH}=\text{AH}$; and if $\beta_1=\beta_2$, then $\text{GH}=\text{AFT}$.

\subsection{The Exponentiated Weibull distribution}\label{sec:ExpWeibull}
We propose modelling the baseline hazard $h_0(t)$ using the Exponentiated Weibull (EW) distribution. The Exponentiated Weibull distribution is simply obtained by exponentiating the Weibull cumulative distribution function to an unspecified positive power \citep{M95}. This simple transformation adds a second shape parameter that, interestingly, induces considerable flexibility to the hazard function. The hazard function of the Exponentiated Weibull distribution can capture several basic shapes: constant, increasing, decreasing, bathtub, and unimodal (See Figure \ref{fig:hazardshapes}), making it appealing for survival models \citep{CM14}. This distribution has been recently used in the context of AFT models \cite{K17}, while alternative families of flexible parametric AFT models have also been studied in \cite{RG16,RH16}. The Exponentiated Weibull probability density and cumulative distribution functions with scale, shape, and power parameters $(\sigma,\kappa,\alpha)$ are given, respectively, by:
\begin{eqnarray}\label{EWPDFCDF}
f_{\text{EW}}(t;\sigma,\kappa,\alpha) &=&  \alpha \dfrac{\kappa}{\sigma} \left(\dfrac{t}{\sigma}\right)^{\kappa-1} \left[1-\exp\left\{-\left(\dfrac{t}{\sigma}\right)^{\kappa}\right\}\right]^{\alpha-1} \exp\left\{-\left(\dfrac{t}{\sigma}\right)^{\kappa}\right\}, \nonumber\\
F_{\text{EW}}(t;\sigma,\kappa,\alpha) &=& \left[1-\exp\left\{-\left(\dfrac{t}{\sigma}\right)^{\kappa}\right\}\right]^{\alpha},
\end{eqnarray}
where $t,\sigma,\kappa,\alpha >0 $. This distribution reduces to the Weibull distribution for $\alpha=1$. The corresponding hazard function is obtained, by definition, as $h_{\text{EW}}(t;\sigma,\kappa,\alpha) = f_{\text{EW}}(t;\sigma,\kappa,\alpha)/\left[1-F_{\text{EW}}(t;\sigma,\kappa,\alpha)\right]$. The EW distribution is identifiable (See Proposition 1 from the Appendix), and general results about the identifiability of the different hazard-based models \eqref{PH}--\eqref{GH} are presented in \cite{C01}. It was shown that the GH model is identifiable except for the case when the baseline hazard is Weibull \cite{C01}. This is not an issue since, as already mentioned, it corresponds to the case when the AFT, PH, and AH models coincide. Moreover, it has also been shown that the AFT and AH models are not identifiable when the baseline hazard is flat \cite{C01} (\textit{i.e.}~exponential), which corresponds to a case when the shape of the baseline hazard does not change with time, a case of little interest in practice.

\subsection{Parameter interpretation}\label{sec:ParInterpret}
Some clarifications on the interpretation of the parameters estimated from the AH and GH models seem appropriate as these models are not as well known as PH and AFT models. Notice that these interpretations directly translate to the HH model as this model is a special case of the GH model.

We start by interpreting the parameters for the AH model \eqref{AH}, as this will facilitate the interpretation for the GH model. The parameter interpretation depends on the shape of the baseline hazard, which we classify here as monotone (increasing/decreasing) or not (bathtub/unimodal).

For monotonic baseline hazard, a positive value of $\beta$ for one unit change in the covariate $x$ means that $x$ has a harmful (beneficial) effect if the baseline hazard is increasing (decreasing) (see panels (a) and (b) in Figures S3--S4. A negative value of $\beta$ means that $x$ has a beneficial (harmful) effect if the baseline hazard is increasing (decreasing) (see panels (a) and (b) in Figures S1--S2). In other words, a positive value of $\beta$ accelerates the progression of the hazard, which is beneficial for the patients if the hazard decreases and harmful if it increases. On the contrary, a negative value of  $\beta$ decelerates the progression of the hazard, which is beneficial for the patients if the hazard increases and harmful if it decreases.

If the shape of the baseline hazard is unimodal (or bathtub), a positive value of $\beta$ accelerates the evolution of the hazard, thus the maximum (minimum) will be reached sooner (see panels (c) and (d) in Figures S3--S4). A negative value of $\beta$ decelerates the evolution of the hazard, thus the maximum (minimum) will be reached later (see panels (c) and (d) in Figures S1--S2).

\begin{figure}[!htbp]
	\begin{center}
		\begin{tabular}{c}
			\includegraphics[width=10cm]{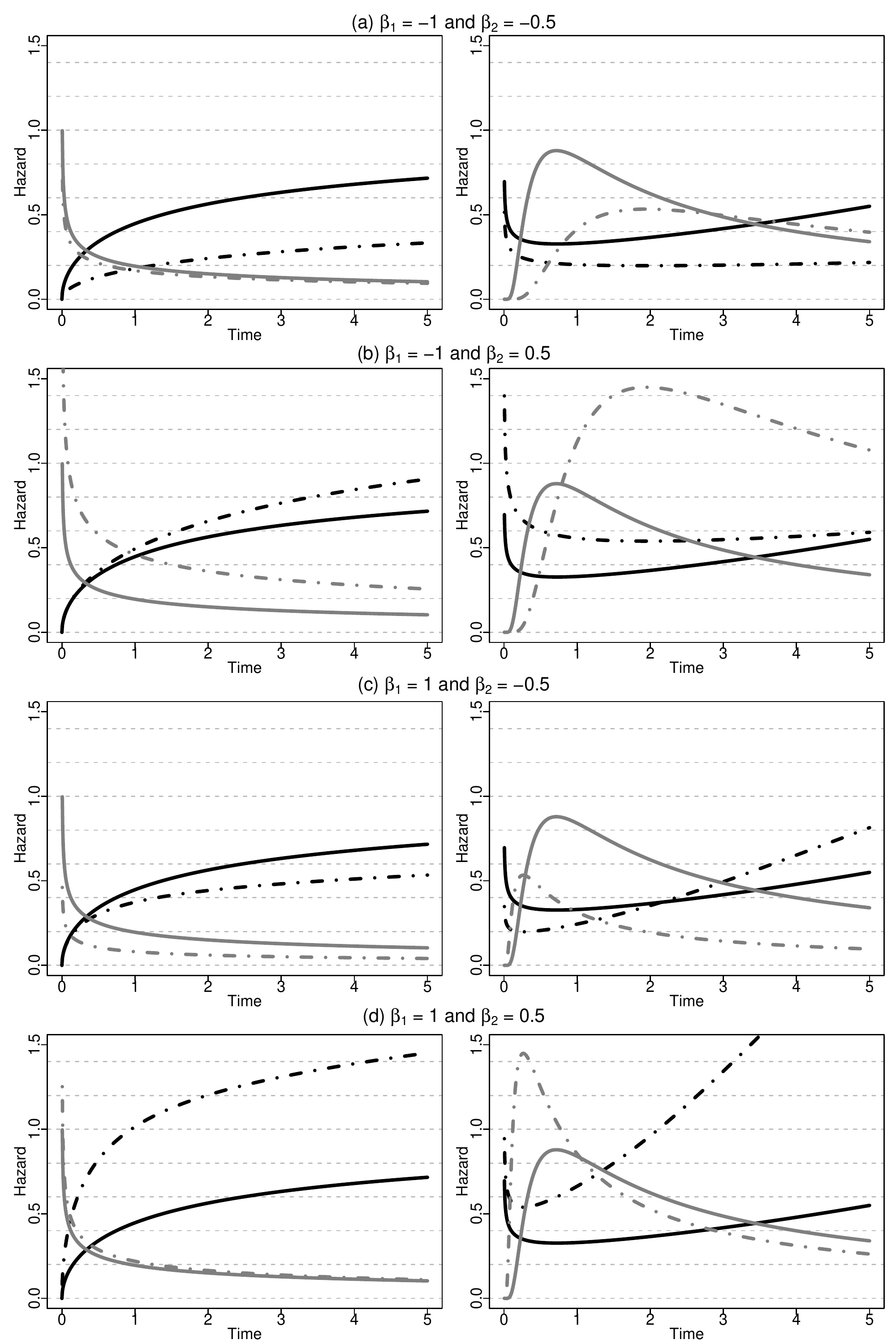}
		\end{tabular}
	\end{center}
	\caption{Hazard shapes obtained with different values of the parameters defining the Exponentiated Weibull distribution $\left(\sigma,\kappa,\alpha\right)$, and different values of the regression coefficients  $\beta_1$ and $\beta_2$ for a binary covariate $x$. Monotonic hazard (increasing in black and decreasing in grey) are displayed on the left column, and bathtub or unimodal (black or grey, respectively) on the right column, where solid lines represent the baseline hazard (\textit{e.g.}~unexposed $x=0$) and dot-dashed lines represent the hazard for the exposed ones ($x=1$). The values of $\left(\sigma,\kappa,\alpha\right)$ are: $(2, 1.2, 1.25)$, $(3.8, 0.5, 1.5)$, $(4.5, 2, 0.4)$, $(0.0002, 0.21, 300)$ for monotonic increasing, monotonic decreasing, bathtub and unimodal shapes, respectively.}
	\label{fig:hazardshapes}
\end{figure}

From the AH model, it is worth noticing that the general shape will not change (a unimodal shape will remain unimodal) and that the peak/minimum (if any) reached by the hazards defined by different subgroups will be at the same level. On the other hand, when using the GH model \eqref{GH} the parameter $\beta_1$ is directly multiplying the time $t$, thus re-scaling the timescale (accelerating or decelerating, \textit{i.e.}~they play a role on the x-axis, changing the pace of the hazard's progression), while the parameter $\beta_2$ is modifying the level of the hazard (role on the y-axis, changing the magnitude of the hazard, Figure \ref{fig:hazardshapes}). This can be clearly seen in the right panels of Figure \ref{fig:hazardshapes} for the unimodal shape: in panels (a) and (b) $\beta_1$ is negative, thus the peak is reached later for patients' group with $x=1$ (dot-dashed grey lines) compared to the baseline group (solid grey lines), while the parameter $\beta_2$ changes the magnitude of the hazard (thus the level of the peak). In panels (c) and (d), $\beta_1$ is positive, thus the peak is reached sooner for patients' group with $x=1$ compared to the reference group, while the peak level is changed according to $\beta_2$. The same interpretation applies for bathtub hazards (black lines in the right panels in Figure \ref{fig:hazardshapes}), and for monotonic hazards (left panels in Figure \ref{fig:hazardshapes}), even though for these later the interplay of both parameters $\beta_1$ and  $\beta_2$ is less obvious to see on the graphs.

The AH and GH models allow a crossover of the hazards (and also of the survival functions). In some cases, this advantage may lead to difficulties in interpreting clearly the parameters \citep{CW00}. Thus, plotting the hazards according to different covariate patterns is recommended to help clarifying the time-to-event process compared to reporting only the survival functions. Indeed, the survival probability, being a cumulative measure, does not help visualising particular features of the instantaneous process \citep{H14}.

\subsection{The likelihood function}\label{sec:Lik}
Let $(t_j,\bx_j,\delta_j)$, $j=1,\dots,n$ be a sample of times to event from a population of cancer patients $t_j>0$, with covariates $\bx_j\in {\mathbb R}^p$, and vital status indicators $\delta_j$ (1-death, 0-censored). Let also $h_{\text{P}}(\text{age}_j+t_j; \text{year}_j+t_j;\bz_j)$ be the corresponding population hazard rates obtained from the national life tables based on the variables $\bz_j \in {\mathbb R}^q$, $q\leq p$, where $\text{age}_j$ represents the age at diagnosis and $\text{year}_j$ the year of diagnosis of patient $j$. The likelihood function of the full vector of parameters $\bm{\psi}$ is then given by
\begin{eqnarray*}
{\mathcal L}_0(\bm{\psi};\text{Data}) &=& \prod_{j=1}^n h(t_j;\bx_j)^{\delta_j}S(t_j;\bx_j),\\
&\propto& \prod_{j=1}^n \left\{ h_{\text{P}}(\text{age}_j+t_j;\text{year}_j+t_j;\bz_j)+ h_{\text{E}}(t_j;\bx_j) \right\}^{\delta_j}\exp\left\{- H_{\text{E}}(t_j;\bx_j)\right\}.
\end{eqnarray*}
Notice that we have removed the term $\exp\left\{- H_{\text{P}}(\text{age}_j+t_j;\text{year}_j+t_j;\bz_j) + H_{\text{P}}(\text{age}_j;\text{year}_j;\bz_j)\right\}$ from the likelihood as it does not depend on the parameters $\bm{\psi}$.

In order to obtain confidence intervals for the parameters, we reparameterise the baseline EW hazard in terms of $(\tilde{\kappa},\tilde{\sigma},\tilde{\alpha}) = (\log\{\kappa\},\log\{\sigma\},\log\{\alpha\})$. Appealing to the consistency and asymptotic normality of the maximum likelihood estimators (MLEs) of the EW distribution, and the availability of large samples in the context of cancer epidemiology, we propose the use of asymptotic confidence intervals of the type $\widehat{\bm{\psi}} \pm  Z_{1-\frac{\tau}{2}} \operatorname{diag}\left(J^{-\frac{1}{2}}(\widehat{\bm{\psi}})\right)$, where $J(\bm{\psi}) = -\dfrac{\partial^2}{\partial \bm{\psi} \partial \bm{\psi}^T} \log {\mathcal L}_0(\bm{\psi};\text{Data})$ is the negative of the Hessian matrix of the log-likelihood function under the appropriate parameterisation, and $1-\tau \in (0,1)$ is the confidence level.

Confidence intervals for the net survival curve at specific time-points are obtained using a simulation-based algorithm \citep{M13}. The idea is to simulate from the asymptotic (multivariate normal) distribution of the parameters in order to obtain a Monte Carlo sample of the net survival at specific time-points, which is used to construct the corresponding confidence intervals.

To measure relative goodness-of-fit of a hazard-based model structure among the ones detailed previously, we employ the Akaike Information Criterion (AIC):
\begin{eqnarray*}
\text{AIC} = -2\log {\mathcal L}_0(\widehat{\bm{\psi}};\text{Data}) + 2k,
\end{eqnarray*}
where $k$ is the number of parameters (the dimension of $\bm{\psi}$), and $\widehat{\bm{\psi}}$ is the MLE of $\bm{\psi}$.

\section{Simulation study}\label{sec:simulation}
In this section, we present an extensive simulation study where we illustrate the good frequentist properties of the proposed models as well as the ability of the AIC criterion to select models that properly capture the underlying hazard structure. The parameter values are chosen in order to produce scenarios that resemble cancer population studies concerning an aggressive type of cancer (relatively low 5-year survival), such as lung cancer.
\subsection{Data generation and simulations designs}
 We simulated $N=1000$ data sets of size $n=1000, 5000, 10000$, assuming the additive hazard decomposition given in \eqref{HazardAddModel}.
The variable ``age'' was simulated using a mixture of uniform distributions with $0.25$ probability on $(30,65)$, $0.35$ probability on $(65,75)$, and $0.40$ probability on $(75,85)$ years old. The binary variables ``sex'' and ``W'' were both simulated from a binomial distribution with probability $0.5$ (the variable ``W'' could be viewed as ``treatment'' or ``comorbidity''). In all the scenarios, we simulated the ``other-causes'' time to event using the UK life tables based on ``age'' and ``sex'', assuming all patients were diagnosed on the same year. The time to event from the excess hazard (cancer event time) was generated using the inverse transform method \cite{Ross2006simulation}, and assuming effects of the 3 variables ``age'', ``sex''  and ``W''. We assumed either (i) only administrative censoring at $T_C=5$ years, which induced approximately $30\%$ censoring in all cases, or (ii) an additional independent random censoring (drop-out) using an exponential distribution with rate parameter $r$. In this latter case, we choose values for $r$ to induce around $50\%$ censoring. We considered 6 generating mechanisms for the excess hazard, all of them with EW baseline hazards: (i) Proportional Hazards (Scenario PH), (ii) Accelerated Failure time (Scenario AFT), (iii) Accelerated Hazard (Scenario AH), (iv) Hybrid Hazard (Scenario HH), (v) General Hazard (Scenario GH), and (vi) a hybrid hazard model where the hazards and the survival curves associated to the variable ``W'' cross (Scenario CH). The values of the model parameters used for each scenario are summarised in Table S1 from the Appendix. Moreover, tables S2--S7 from the Appendix present the 1-year and 5-year net survival implied by the different combinations of hazard structures and covariate values.

\subsection{Analysis of the simulated data}
To assess the frequentist properties of the models in all simulation scenarios, we fitted the GH model with EW baseline hazard (GHEW), as compared to the corresponding true generating model. Additionally, to measure the ability of the AIC to select the true generating model, we also fitted the remaining models in each scenario. In all cases, we used the true parameter values of the generating model as initial points in the optimisation step because our aim was more to study the estimator's properties rather than the properties of the optimisation process. However, as the interest of analysts is also on the overall properties (\textit{i.e.}~including the optimisation process properties in the full process of estimation), we present in the Appendix a thorough study where we present three alternative ``automatic'' choices for the initial points which we will later use in the real data example (see section 5 in the Appendix). We present results about the convergence using these initial points as well as the resulting estimators, which are virtually the same as those obtained with the chosen initial points, as expected, with the expense of a higher computing time.

We performed the analysis using R software. The optimisation step was conducted using the R commands `nlminb()' and `optim()', while the Hessian matrix used in the construction of the asymptotic confidence intervals are approximated using the command `hessian()' from the R package `numDeriv', which represents an efficient method to calculate the Hessian matrix. The cases where the command `hessian()' produced `Inf' or `NaN' values were excluded as this merely represents a numerical problem associated to numerical differentiation, which mainly affects the most difficult cases (\textit{i.e.} with $n=1000$ and $50\%$ censoring).

\subsection{Measures of performance}
We report the mean and median of the MLEs for the corresponding models, as well as the empirical standard deviation, the mean (estimated) standard error, the root-mean-square error, and coverage proportions of asymptotic confidence intervals. In addition, we report the proportion of times the different fitted models are selected using AIC.
The excess hazard functions associated with the best models selected using AIC are plotted in Figures S5--S14, for different covariate patterns, and compared to the true generating model.

\subsection{Simulation results}
\begin{figure}[!htbp]
\begin{center}
\begin{tabular}{c}
\includegraphics[width=12cm, height=10cm]{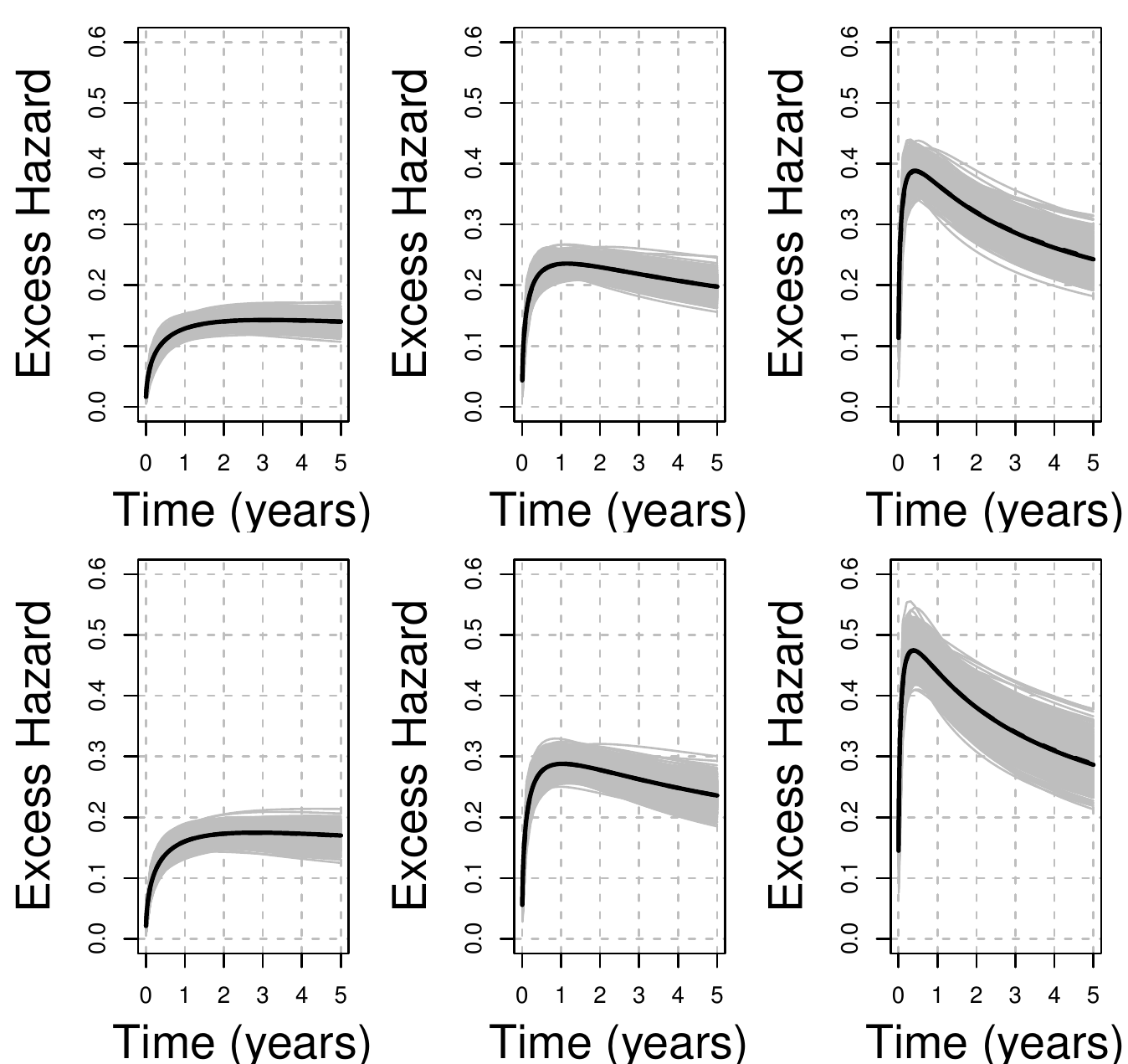}
\end{tabular}
\end{center}
\caption{Mean of the best fitted hazards in terms of AIC (dashed lines), compared to the true generating hazard (continuous lines), and 1000 sample-specific fitted hazards (grey lines) in the scenario GH for $n=5000$ and $30\%$ censoring. Panels from left to right correspond to covariate values (age, sex, comorbidity)=$(60,0,0),(70,0,0),(80,0,0),(60,1,0),(70,1,0),(80,1,0)$, respectively. The dashed and continuous lines are virtually the same.}
\label{fig:GH5000}
\end{figure}

The results are presented in Tables S8--S29 in the Appendix. For illustrative purposes, we present in Tables \ref{table:GH5000}--\ref{table:AICGH} and Figure \ref{fig:GH5000} the results for Scenario GH with $n=5000$. In this scenario, we observe very good performance of our approach, with very small bias and a coverage close to the nominal value of 95\% (Table \ref{table:GH5000} and Figure \ref{fig:GH5000}). Regarding performances of model selection using the AIC, the true model was selected in around 1/3 of the simulated datasets in the worst case (sample size of 1000 and 50\% of censoring), but this proportion increases to 90\% or more in situations with a larger sample size (Table \ref{table:AICGH}). More generally, from this simulation study, we can conclude that the MLEs are close to the true values for moderate samples, even in the case of a high censoring rate (Tables S8, S12, S16, S20, S24, S26) and that the coverage is usually quite close to $95\%$. As expected, the RMSE decreases as the sample size increases (or with identical sample size but lower censoring percentage). For $n=5000$, we observe low bias and variance of the MLEs, and that the AIC selects the correct model with a high proportion. For $n=10000$, the AIC selects the true model in more than $85\%$ of cases for all scenarios and whatever the censoring level. Interestingly, in the GH scenario, as long as the sample size was equal to 5000 or larger, the AIC selected the true model in $90\%$  or more samples, regardless of the censoring level. Moreover, even in  cases where the incorrect model is selected using AIC, we see that the baseline hazard is close to the true generating model, reflecting that the AIC selects the model closest to the true generating model (see figures S5--S14).

\begin{table}[!htbp]
\centering
\begin{tabular}{|cccccccc|}
\hline
Model & Parameter & MMLE & mMLE & ESD & Mean Std Error & RMSE & Coverage \\
\hline
\multicolumn{8}{|c|}{$30\%$ censoring } \\
\hline
\multirow{ 7}{*}{GHEW} & $\sigma$ (1.75) & 1.747 & 1.726 & 0.405 & 0.409 & 0.405 & 0.950 \\
&  $\kappa$ (0.6) & 0.601 & 0.600 & 0.063 & 0.064 & 0.063 & 0.946 \\
&  $\alpha$ (2.5) & 2.576 & 2.523 & 0.461 & 0.459 & 0.467 & 0.947 \\
&  $\beta_{11}$ (0.1) & 0.100 & 0.100 & 0.013 & 0.013 & 0.013 & 0.961 \\
&  $\beta_{12}$ (0.1) & 0.112 & 0.109 & 0.233 & 0.227 & 0.233 & 0.949 \\
&  $\beta_{13}$ (0.1) & 0.099 & 0.104 & 0.230 & 0.228 & 0.230 & 0.956 \\
&  $\beta_{21}$ (0.05) & 0.050 & 0.050 & 0.003 & 0.003 & 0.003 & 0.958 \\
&  $\beta_{22}$ (0.2) & 0.201 & 0.202 & 0.041 & 0.042 & 0.041 & 0.951 \\
&  $\beta_{23}$ (0.25) & 0.251 & 0.251 & 0.043 & 0.042 & 0.043 & 0.943 \\
\hline
\multicolumn{8}{|c|}{$50\%$ censoring } \\
\hline
\multirow{ 7}{*}{GHEW} & $\sigma$ (1.75) & 1.741 & 1.736 & 0.457 & 0.465 & 0.457 & 0.957 \\
&  $\kappa$ (0.6) & 0.600 & 0.600 & 0.074 & 0.076 & 0.074 & 0.948 \\
&  $\alpha$ (2.5) & 2.601 & 2.508 & 0.556 & 0.539 & 0.565 & 0.963 \\
&  $\beta_{11}$ (0.1) & 0.102 & 0.101 & 0.016 & 0.016 & 0.016 & 0.959 \\
&  $\beta_{12}$ (0.1) & 0.090 & 0.095 & 0.267 & 0.266 & 0.267 & 0.957 \\
&  $\beta_{13}$ (0.1) & 0.092 & 0.093 & 0.269 & 0.268 & 0.269 & 0.948 \\
&  $\beta_{21}$ (0.05) & 0.050 & 0.050 & 0.004 & 0.004 & 0.004 & 0.945 \\
&  $\beta_{22}$ (0.2) & 0.201 & 0.200 & 0.045 & 0.047 & 0.045 & 0.956 \\
&  $\beta_{23}$ (0.25) & 0.252 & 0.251 & 0.049 & 0.048 & 0.049 & 0.949 \\
\hline
\end{tabular}
\caption{Simulation results for the scenario GH with $(\sigma,\kappa,\alpha) = (1.75,0.6,2.5)$, $\beta_1 = (0.1,0.1,0.1)$, $\beta_2 = (0.05,0.2,0.25)$, and $n=5000$. Mean of the MLEs (MMLE), median of the MLEs (mMLE), empirical standard deviation (ESD), mean (estimated) standard error, root-mean-square error (RMSE), and coverage proportions (Coverage).}
\label{table:GH5000}
\end{table}

\begin{table}[!htbp]
\centering
\begin{tabular}{|ccc|}
\hline
Model &  $30\%$ censoring & $50\%$ censoring \\
\hline
\multicolumn{3}{|c|}{$n=1000$ } \\
\hline
PHEW  & 1.6  &  2.7   \\
AHEW  &  0 &    0 \\
AFTEW &  56.5 &  64.5   \\
GHEW  &  41.9 &    32.8  \\
\hline
\multicolumn{3}{|c|}{$n=5000$ } \\
\hline
PHEW  & 0  &   0  \\
AHEW  & 0  &    0 \\
AFTEW & 3.2  &   10  \\
GHEW  & 96.8  &  90    \\
\hline
\multicolumn{3}{|c|}{$n=10000$ } \\
\hline
PHEW  &  0 &   0  \\
AHEW  &  0 &   0  \\
AFTEW &  0 &   0.5  \\
GHEW  & 100  &   99.95   \\
\hline
\end{tabular}
\caption{Percentage of models selected with AIC in the scenario GH with $(\sigma,\kappa,\alpha) = (1.75,0.6,2.5)$, $\beta_1 = (0.1,0.1,0.1)$, and $\beta_2 = (0.05,0.2,0.25)$.}
\label{table:AICGH}
\end{table}

\section{Real data example}\label{sec:Example}

To illustrate the new proposed models, we analysed a dataset obtained from population-based national cancer registry of lung cancer patients diagnosed in 2012 in the United-Kingdom. We linked these data to administrative data (Hospital Episode Statistics -HES- and Lung Cancer Audit data -LUCADA-) and applied specific algorithms to derive information on stage at diagnosis and presence of comorbidities at the time of diagnosis \cite{benitez2016, maringe2017}. To retrieve information on comorbidity, we used a 6-year period up to 6 months before diagnosis where we checked for the presence of any of 18 comorbidities that are used to define the Charlson Comorbidity Index (CCI), in addition to obesity \cite{maringe2017}. The information was then dichotomised into 2 categories: ``no comorbidity'' \textit{vs.} ``at least 1 comorbidity (comorbidity indicator = 1)'' in our illustrative example. We measured deprivation using the Income Domain from the 2010 England Indices of Multiple Deprivation, defined at the Lower Super Output Area level (mean population $1500$). The Income Domain measures the proportion of the population in an area experiencing deprivation related to low income, and ranges from $0$ to $77\%$ (\url{https://www.gov.uk/government/statistics/english-indices-of-deprivation-2010}). Follow-up was assessed on the 31st of December 2015, at which time patients alived were censored (so the maximum follow-up was 4 years). We restricted our analysis to women with no missing data, and applied the PH model \eqref{PH}, the AH model \eqref{AH}, the AFT model \eqref{AFT} and the GH model \eqref{GH}, with an Exponentiated Weibull distribution for the baseline hazard. The variables included in the models are `agediagc' (centred age at diagnosis), `Istage2', `Istage3', and `Istage4' (stage at diagnosis), `INCOME\_SCORE\_2015c' (centred Income Domain), `comorbidity' (presence of comorbidity). The time dependent effects are indicated in Table \ref{table:realdataMLE} with the subindex `$t$'.

\begin{figure}
\begin{center}
\begin{tabular}{c}
\includegraphics[width=12cm, height=8cm]{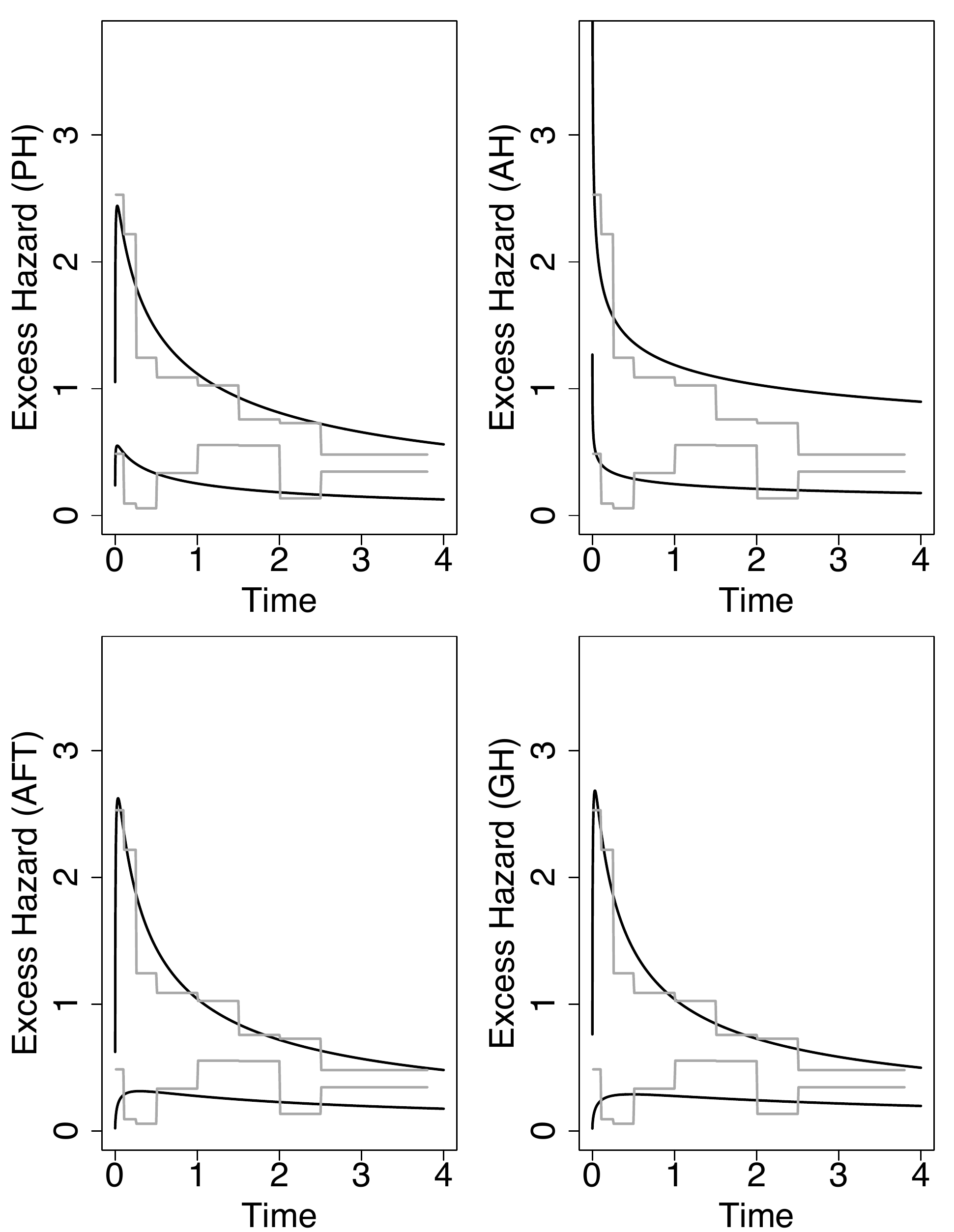} \\
(a)\\
\includegraphics[width=12cm, height=8cm]{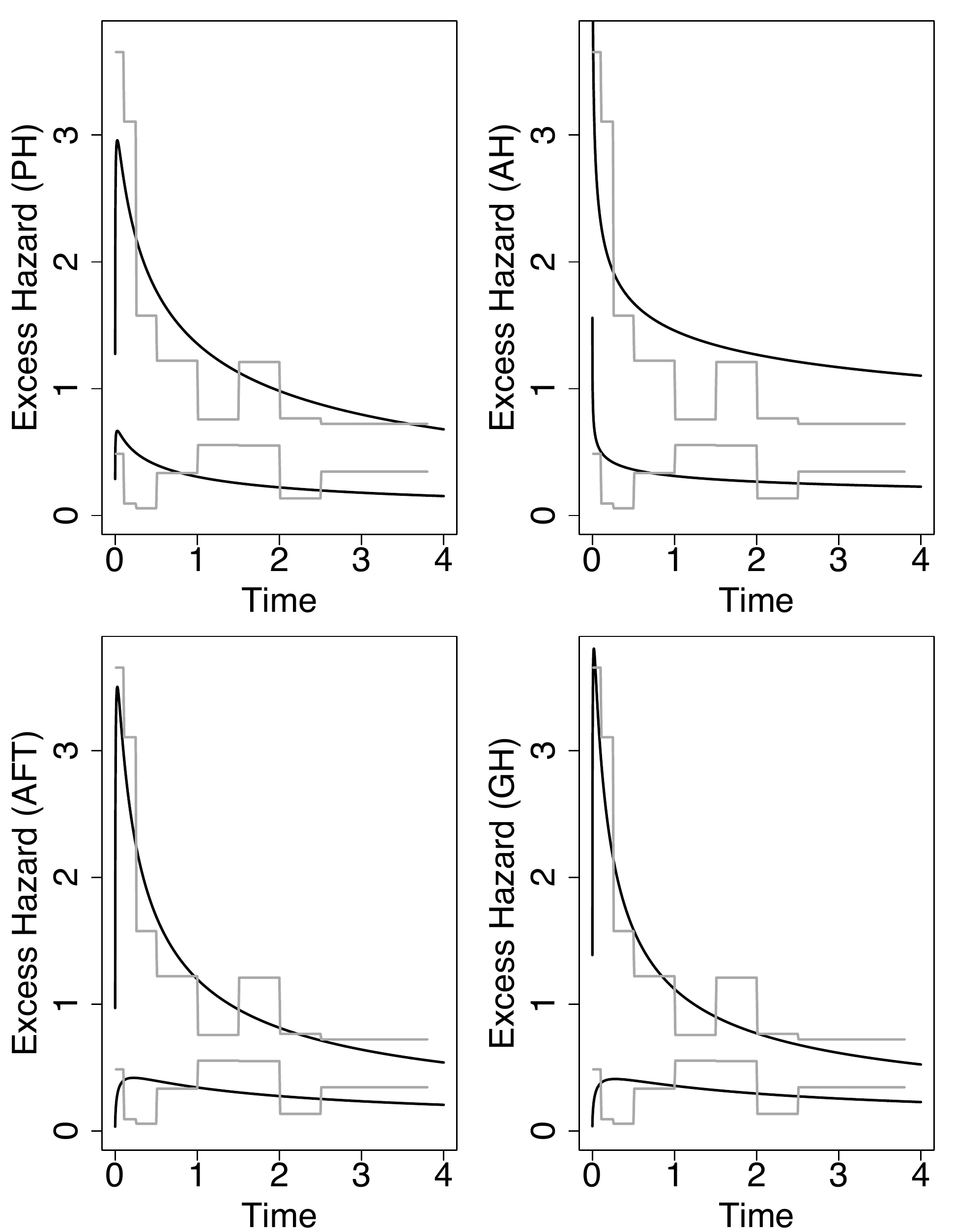} \\
(b)
\end{tabular}
\end{center}
\caption{Shapes of the EW excess hazard for PH, AFT, AH, and GH models with baseline covariate values (solid lines) \textit{vs.}~Piecewise excess hazard (grey line): (a) comorbidity = 0 and Stage II (the 2 lowest curves/step functions in each cell), comorbidity = 0 and Stage IV (the 2 highest curves/step functions in each cell); and (b) comorbidity = 1 and Stage II (the 2 lowest curves/step functions in each cell), comorbidity = 1 and Stage IV (the 2 highest curves/step functions in each cell).}
\label{fig:PWC01}
\end{figure}

We observed $n=14557$ patients with complete cases among which $n_o=12138$ died before the 31st of December 2015. The median follow-up among patients censored was $3.46$ years. The $25\%$, $50\%$ and $75\%$ quantiles of the patients' age at diagnosis was $64.9$, $72.6$, $80.2$ while the mean was $72.0$. Among the patients, $2434$ were Stage I, $1131$ were Stage II, $3421$ were Stage III, and $7751$ were Stage IV. Finally, $4318$ patients were classified with comorbidity indicator 1. Results of this analysis are presented in Table \ref{table:realdataMLE}. From this table, we observe that the GHEW is clearly favoured by the AIC (followed by the AFTEW model), thus suggesting the need for including time-dependent as well as proportional effects. The signs of all the estimates are positive in this model. This implies that an increase of one unit in the value of any of the covariates leads to an acceleration of the time to reach the maximum of the hazard as well as an increase of the maximum of the hazard. The magnitude of such acceleration is given by the value of the corresponding estimated parameter, and we noticed that these two effects are different for each variable thus explaining the better fit of the GH model compared to the AFT model. We have compared the GHEW model against alternative models with fewer covariates but they were not favoured by the AIC. For comparison, we have also used the `mexhaz' R package \cite{C16, MexhazPackage} to produce models with time-dependent effects (using the command {\tt nph}) for some or all the variables and we have found that none of these models provided a better fit than the GHEW in terms of AIC (with `mexhaz', the model with the lowest AIC ($=20270.44$) was the model assuming time-dependent effects for all variables). Figures \ref{fig:PWC01}a and \ref{fig:PWC01}b show the shapes of the hazard functions associated to the different models for patients aged 70 years at diagnosis, an income score value of 0.15 and with either a stage II or a stage IV cancer, and without comorbidity (panel (a)) or with comorbidity (panel (b)). These figures also show the piecewise excess hazard estimated separately on each of the 4 groups defined by the following values of covariate: $\text{age}\in[65,75]$, $\text{Income Score} \in [0.1,0.2]$, Stage II or Stage IV, and $\text{Comorbidity}=0 \text{ or } 1$. Those piecewise hazards represent a way to check the quality of the fit of the different models \citep{R07}. These figures suggest a good fit of the GH model overall, while the excess hazards obtained from the AH model are always over-estimated after 6 months for the group of stage IV patients. For the  group of stage IV patients with comorbidity (panel (b)), the PH model does not capture the high excess hazard just after the diagnosis and the sharp decrease that follows.


Table \ref{table:realdataNS} presents the net survival at $1,2,3,3.9$ years, for the total population (Comorbidity = 0, 1) and for a subgroup (age-group 55-65, Stage I, and Comorbidity = 0, 1, ), using the GHEW model and the Pohar-Perme nonparametric estimator \cite{PS12}. The confidence intervals for the net survival in the GHEW model were obtained using the simulation-based algorithm described in Section \ref{sec:Lik} using $B=10,000$ samples from the asymptotic distribution of the parameters. The Pohar-Perme estimator and its corresponding confidence intervals were calculated using the `relsurv' R package. We observe that the results with the parametric and nonparametric approaches are very close, and that the confidence intervals for the parametric model are slightly shorter (as expected), which shows that the proposed parametric model can accurately capture the underlying hazard structure.

\begin{table}
\centering
\begin{tabular}{|rcccc|}
  \hline
Model & PHEW & AHEW & AFTEW & GHEW \\
  \hline
  scale                  & 0.059 (0.038) & 8.482   (0.724) & 1.190 (0.175) & 1.838 (0.374) \\
  shape                  & 0.188 (0.014) & 0.539   (0.046) & 0.385 (0.012) & 0.442 (0.033) \\
  power                  & 9.175 (1.420) & 1.483   (0.129) & 4.387 (0.312) & 3.593 (0.368) \\
  $\text{agediagc}_t$             & --            & -0.112  (0.006) & 0.032 (0.001)            & 0.041 (0.004) \\
  $\text{Istage2}_t$             & --            & -2.977  (0.282) &  0.881 (0.065)           & 0.691 (0.311) \\
  $\text{Istage3}_t$             & --            & -6.680  (0.337) &  1.909 (0.050)           & 1.707 (0.229) \\
  $\text{Istage4}_t$             & --            & -10.469 (0.416) &  3.003 (0.046)           & 3.413 (0.226) \\
  $\text{INCOME\_SCORE\_2015c}_t$ & --            & -2.668  (0.416) &  0.744 (0.115)           & 0.822 (0.448) \\
  $\text{comorbidity}_t$          & --            & -1.021  (0.106) &  0.289 (0.029)           & 0.539 (0.114) \\
  agediagc               & 0.022 (0.001) & --              & -- ($\star$) & 0.034 (0.001) \\
  Istage2                & 0.721 (0.056) & --              & -- ($\star$) & 0.845 (0.069) \\
  Istage3                & 1.473 (0.043) & --              & -- ($\star$) & 1.849 (0.053) \\
  Istage4                & 2.211 (0.041) & --              & -- ($\star$) & 3.073 (0.050) \\
  INCOME\_SCORE\_2015c   & 0.527 (0.085) & --              & -- ($\star$) & 0.750 (0.150) \\
  comorbidity            & 0.192 (0.021) & --              & -- ($\star$) & 0.349 (0.039) \\
  AIC                    & 20523.141 & 20855.753 & 20189.124 & \bf{20164.911}  \\
   \hline
\end{tabular}
\caption{Maximum likelihood estimates of the parameters (standard errors) for the different excess hazard models, with their corresponding AIC.  The time dependent effects are indicated with the subindex `$t$'. ($\star$) By construction, the effects of covariates are constrained to be the same for the time-dependent and time fixed effects in the AFT model \eqref{AFT}. See equations \eqref{PH}-\eqref{GH} for more details on the different hazard structures.}
\label{table:realdataMLE}
\end{table}


\begin{table}
\centering
\begin{tabular}{|cccccccc|}
\hline
\multicolumn{8}{|c|}{\underline{Total population by comorbidity}} \\
&&&&&&&\\
& & \multicolumn{3}{c}{GHEW} &  \multicolumn{3}{c|}{Pohar-Perme} \\
&&&&&&&\\
Comorb. & year & NS & lower & upper & NS & lower & upper \\
&&&&&&&\\
\multirow{4}{*}{0}   & 1   & 0.407 & 0.401 & 0.416 & 0.408 & 0.398 & 0.418 \\
                     & 2   & 0.270 & 0.265 & 0.278 & 0.268 & 0.259 & 0.277 \\
                     & 3   & 0.204 & 0.199 & 0.211 & 0.209 & 0.201 & 0.218 \\
                     & 3.9 & 0.167 & 0.162 & 0.174 & 0.182 & 0.172 & 0.191 \\
&&&&&&&\\
\multirow{4}{*}{1}   & 1   & 0.380 & 0.371 & 0.391 & 0.370 & 0.356 & 0.385 \\
                     & 2   & 0.254 & 0.246 & 0.264 & 0.238 & 0.225 & 0.252 \\
                     & 3   & 0.193 & 0.185 & 0.203 & 0.169 & 0.158 & 0.182 \\
                     & 3.9 & 0.158 & 0.150 & 0.168 & 0.138 & 0.126 & 0.152 \\
&&&&&&&\\
\multicolumn{8}{|c|}{\underline{Age group 55-65 at Stage I by comorbidity}} \\
&&&&&&&\\
Comorb. & year & NS & lower & upper & NS & lower & upper \\
&&&&&&&\\
\multirow{4}{*}{0}  & 1   & 0.924 & 0.914 & 0.935 & 0.928 & 0.897 & 0.960 \\
                    & 2   & 0.842 & 0.827 & 0.860 & 0.837 & 0.794 & 0.883 \\
                    & 3   & 0.770 & 0.752 & 0.791 & 0.797 & 0.749 & 0.847 \\
                    & 3.9 & 0.712 & 0.692 & 0.737 & 0.762 & 0.705 & 0.823 \\
&&&&&&&\\
\multirow{4}{*}{1}  & 1   & 0.881 & 0.869 & 0.896 & 0.901 & 0.851 & 0.953 \\
                    & 2   & 0.772 & 0.754 & 0.793 & 0.744 & 0.674 & 0.821 \\
                    & 3   & 0.684 & 0.662 & 0.710 & 0.670 & 0.595 & 0.755 \\
                    & 3.9 & 0.618 & 0.595 & 0.647 & 0.627 & 0.541 & 0.725 \\
\hline
\end{tabular}
\caption{Net survival (NS) at $T=1,2,3,3.9$ years and $95\%$ confidence intervals, estimated using the GHEW model and the non parametric Pohar-Perme estimator.}
\label{table:realdataNS}
\end{table}

\newpage

\section{Discussion}\label{sec:discussion}

We have studied a general parametric hazard structure that can capture the basic shapes of the baseline hazard and that contains, as particular cases, the main models of interest in survival analysis: proportional hazards, accelerated failure time, and accelerated hazards models. PH and AFT models already enjoy popularity in survival analysis. However, the limited application of the AH model is mainly due to the lack of efficient and reliable estimation methods \cite{C14}. We have shown that, by assuming a flexible parametric distribution function such as the EW, it is possible to conduct classical likelihood inference using already available optimisation algorithms. The combination of such flexible parametric hazard function with the GH structure represents a powerful tool for modelling survival times. Although we have focused on the context of \textit{excess hazard models}, the proposed flexible parametric GH model is also applicable in the context of \textit{overall survival}. We have employed the EW distribution for modelling the baseline hazard in the GH model \eqref{GH}, however, there exist other flexible parametric distributions, such as the generalised Gamma \citep{CM14} and the generalised Weibull \citep{M96} distributions, that can also capture the basic shapes of the hazard function. Here, we only employed the EW distribution as this choice allows for a parsimonious implementation, facilitates the interpretation of the parameters, and the corresponding maximum likelihood estimators (MLEs) are consistent and asymptotically normal in the presence of censored observations \citep{Q12}. The simulation study shows that the proposed model has good frequentist properties and that the selection of an appropriate model structure is feasible for large enough samples. The proposed model can also capture cases where the hazard and the survival curves cross, a case of great interest in practice. Cases when the AH and AFT structures produce crossing hazards have been studied in \cite{Z09}, and these are also illustrated in our simulation study. Despite the flexibility of the EW distribution, there still exist baseline hazard shapes that cannot be captured by this model (such as multimodal hazards). We have conducted additional simulation studies in order to assess the effect of model misspecification on the estimation of net survival. We found that in cases where the EW distribution cannot capture the true shape of the baseline hazard, the net survival functions associated to the fitted models tend to be relatively close to the true model, and that the AIC selects the correct hazard structure with high proportion, provided that the departures from the shapes the EW can capture are moderate. If the departures are severe, this may, unsurprisingly, affect the selection of the correct hazard structure but the selected model tend to resemble the shape of the true generating model (see \citealp{H92} for a general study of fitting parametric survival models under possible misspecification). These additional studies are available from the authors' websites, as well as the R code for fitting the models detailed in this paper.

Although we have centered our attention on the GH structure \eqref{GH}, we point out the more general representation discussed in \cite{C03}:
\begin{eqnarray*}
h_{\text{E}}\left(t;\bx_j\right) = \Lambda\left( h_0(t), \bx_j^T\beta \right),
\end{eqnarray*}
where $\Lambda(\cdot)$ is a known non-negative function that defines the relationship between the baseline hazard and the covariates. This representation contains, for instance, the additive hazards model \citep{LY01} $h_{\text{E}}^{\text{AD}}\left(t;\bx_j\right) = h_0(t) + \bx_j^T\beta$, which we do not consider here as it requires additional conditions to guarantee that $h_{\text{E}}^{\text{AD}}>0$; as well as structures including non-linear relationships. We also point out that the use of splines for including non-linear effects can be coupled with any of the aforementioned hazard structures. A potentially useful choice are B-splines, which are implemented in R.

We have employed the AIC for selecting the hazard structure that better fits the data since this criterion can account for possible model misspecification, as this tool asymptotically selects the model that minimises the Kullback-Leibler divergence between the fitted models and the true generating model, under mild regularity conditions. We point out that other criteria such as the Bayesian Information Criterion (BIC) or cross validation can be employed instead. An interesting feature of the GH model is that, by selecting the active variables, we automatically select the hazard structure that better fits the data. The study of efficient variable selection methods in the general structure \eqref{GH} would allow the identification of active variables as well as the underlying structure (PH, AH, HH, AFT, or GH) of the hazard function. This points out an interesting research line. The simulation study also illustrates the importance of accounting for sparsity, as the inclusion of spurious variables may bias the estimates of the active variables and the parameters of the baseline hazard. Since, in recent years, more variables such as comorbidities and types of treatments have become available at the population level, we believe this topic will become relevant in cancer epidemiology.

\section*{Acknowledgements }

\textbf{Funding:} This research was partly supported by Cancer Research UK grant number C7923/A18525. The findings and conclusions in this report are those of the authors and do not necessarily represent the views of Cancer Research UK.

\textbf{Ethical approvals:} We obtained the ethical and statutory approvals required for this research (PIAG 1-05(c)/2007; ECC 1-05(a)/2010); ethical approval updated 6 April 2017 (REC 13/LO/0610). We attest that we have obtained appropriate permissions and paid any required fees for use of copyright protected materials.

\bibliographystyle{plainnat}
\bibliography{references}

\end{document}